\documentclass[12pt]{iopart}
\usepackage{iopams}
\usepackage{setstack}
\usepackage{lmodern} 
\usepackage{graphicx}
\usepackage{epigraph}
\usepackage{amssymb}
\usepackage{paralist}
\usepackage{color}
\usepackage{xfrac} 

\usepackage{amscd}
\usepackage{dsfont}
\usepackage{bm}

\def\H{ {\mathcal{H}}}

\def\Im{ {\mathrm{Im}}}
\def\Re{ {\mathrm{Re}}}

\newif\ifdraft
\draftfalse

\DeclareGraphicsExtensions{.pdf, .png, .bmp, .jpg}

\renewcommand{\Re}{\operatorname{Re}}

\newcommand{\operatorname}[1]{\text{#1}}
\newcommand{\eqref}[1]{(\ref{#1})}

\begin{document}
\title[Trace formula for spin chains]{Trace formula for spin chains}

\author{Daniel Waltner, Petr Braun, Maram Akila, Thomas Guhr}
\address{
Faculty of Physics, University Duisburg-Essen, Lotharstr. 1, 47048 Duisburg, Germany;
}
\begin{abstract}
While detailed information about the semiclassics for single-particle systems is available, much less is known about the connection between quantum and classical
dynamics for many-body systems. As an example, we focus on spin chains which are of considerable conceptual and practical importance.
 We derive a trace formula for coupled spin $j$ particles which relates the quantum energy levels to the classical dynamics. 
 Our derivation is valid in the limit $j\rightarrow\infty$ with $j\hbar={\rm const.}$ and applies to time-continuous as well as to periodically driven dynamics.
 We provide a simple explanation why the Solari-Kochetov phase can be omitted if the correct classical Hamiltonian is chosen.
\end{abstract}


Keywords: Spin chain, trace formula, coherent states, Solari-Kochetov phase
\maketitle
\section{Introduction}

Connecting quantum properties such as the energy spectral distribution to the
classical properties such as the periodic orbits is a central issue in the
theory of quantum chaos. A milestone was here the derivation of the Gutzwiller
trace formula \cite{Gutz} for chaotic one-particle-systems in 1960s. Besides
connecting classical dynamics and quantum properties for simple model systems
with mixed dynamics such as the hydrogen atom in a strong magnetic field
\cite{Holle}, it also provided analytical understanding \cite{Muller} of the
applicability of Random Matrix Theory \cite{Guhr} to  quantum systems with
classically fully chaotic counterpart and thereby strongly corroborates the 
Bohigas-Giannoni-Schmit conjecture \cite{Bohigas}.

Nowadays, the research focus switches more and more to interacting many
particle systems. In this context, the trace formula for  many particle
systems consisting of indistinguishable particles was derived in \cite{Weide},
the trace formula for Bose-Hubbard Models was obtained in \cite{Engl}. A
prominent many-particle system on which we focus in this paper, are spin chains.
The existing investigations refer to the spin quantum number
$j=1/2$ \cite{Lieb,Atas,Keating,Akila,Simon,Kim} and to larger spin
quantum number both on the side of the theorists \cite{Garg0,Garg,Braun1,Ribeiro1,Ribeiro} as well as
experimentalists \cite{Scazza}.

We derive here a trace formula for the spin chains expressing their spectral
density in terms of the classical periodic orbits. This formula is
asymptotically valid in the limit of the spin quantum number $j\rightarrow\infty$, where $j\hbar$ is kept constant.

The investigations of the semiclassical spin evolution are usually performed
in the basis of the spin coherent states, the motivation  being that the angle
eigenbasis does not provide states with well localized values of $j$.
Therefore the only alternative would be to use as the basis the eigenstates of
the $z$-component of the angular momentum which would lead to the discrete
semiclassics \cite{Braun,Hemmen,Villain,Garg1}. A single spin is
semiclassically a system with one degree of freedom such that the appropriate
quantization condition is the Bohr-Sommerfeld rule; it is given in the
coherent state basis in \cite{Garg2}. In Ref.\ \cite{Ribeiro} a trace formula was derived
in the coherent state basis for one particle with an orbital and one spin
degree of freedom with the energy dependence on the spin variable disregarded.
We, however, consider systems of an arbitrary number of interacting spins where the
trace formula is the adequate tool of investigation.

\bigskip We start in section \ref{secII} with the time dependent propagator
in the coherent state basis as derived in Ref.\ \cite{Garg} and reformulated  in
terms of an intuitively obvious classical Hamiltonian. We give an elementary
proof that this Hamiltonian is the correct one. Thus the
Solari-Kochetov corrections \cite{Sola} that attracted a lot of attention in the last years \cite{Garg0,Garg,Braun1} are not needed provided one observes the
elementary rules of semiclassics such as replacing $\sqrt{j(j+1)}$ with $j+1/2$ 
instead of $j$. In section \ref{secIII} we compute the trace of the
propagator and finally perform the Fourier transform from time to 
energy domain to obtain the spectral density. We note that these steps are
performed in an order different from the derivation of the famous Gutzwiller
trace formula \cite{Gutz}.  In section \ref{secIV} we discuss how these results can
be generalized to periodically driven systems. We conclude in section \ref{secV}. Technical
details are relegated to appendices.

\section{Propagator}\label{secII}
This section is devoted to the propagator for a spin chain consisting of $N$ spins.
In subsection \ref{subsec21} we recapitulate the known expression for the propagator in the limit $j\rightarrow\infty$. 
In the subsection \ref{subsec22} we show that the Solari-Kochetov phase is not needed if the classical 
Hamiltonian is chosen in an appropriate way.
\subsection{Semiclassical expression for the propagator}\label{subsec21}

The propagator for a system of $N$ spins in the coherent state
basis is derived in \cite{Garg} as a generalization of the propagator for
one spin \cite{Garg0}. The Hamiltonian $\mathcal{H}=\mathcal{H}({\boldsymbol{\mathcal{J}}}%
_{1},\ldots,{\boldsymbol{\mathcal{J}}}_{N})$ with ${\boldsymbol{\mathcal{J}}}_{i}=(\mathcal
{J}_{i,1},\mathcal{J}_{i,2},\mathcal{J}_{i,3})$ is a Hermitian polynomial in the
Cartesian components of the spin operators $\boldsymbol{\mathcal{J}}_{i}$,
$i=1,\ldots,N$ with real coefficients. This is no restriction as every Hamiltonian can be expressed like that 
using the commutation relations for the spin operators.
 The matrix element of the propagator between the initial state
$|\mathbf{U}^{\prime}\rangle$ and the final state $|\mathbf{V}^{\prime\prime
}{}^{\ast}\rangle$ where the star denotes complex conjugation, is given by
\begin{equation}
K(\mathbf{U}^{\prime},\mathbf{V}^{\prime\prime},t)=\left\langle \mathbf{V}%
^{\prime\prime}{}^{\ast}\left\vert \mathrm{e}^{-i\mathcal{H}t/\hbar
}\right\vert \mathbf{U}^{\prime}\right\rangle .
\end{equation}
We use here the convention that the primed variables refer to initial and
double primed variables to final coordinates. 
Here $\left\vert \mathbf{U}^{\prime}\right\rangle $ and $\left\vert
\mathbf{V}^{\prime\prime}\right\rangle $ are the direct products of
unnormalized single particle coherent states
\begin{equation}
\left\vert \mathbf{U}^{\prime}\right\rangle =\bigotimes_{i=1}^{N}\left\vert
U_{i}^{\prime}\right\rangle ,\hspace*{4mm}\left\vert \mathbf{V}^{\prime\prime
}\right\rangle =\bigotimes_{i=1}^{N}\left\vert {V^{\prime\prime}}%
_{i}\right\rangle,
\end{equation}
where
\begin{equation}
\left\vert U_{i}^{\prime}\right\rangle =\mathrm{e}^{U_{i}^{\prime}\mathcal
{J}_{i,+}/\hbar}\left\vert j,-j\right\rangle _{i} \label{cohstate}%
\end{equation}
with $U_{i}^{\prime}$ any complex number, $\mathcal{J}_{i,+}=\mathcal{J}_{i,1}%
+i\mathcal{J}_{i,2}$ the spin raising operator and $\left\vert j,-j\right\rangle
_{i}$ the lowest eigenstate of $\mathcal{J}_{i,3}$ and $\boldsymbol{\mathcal{J}}_i^2$ with magnetic 
quantum number $-j$.
Accordingly, $\left\vert U_{i}^{\prime}\right\rangle $ is an eigenstate with
the eigenvalue $\hbar j$ of the angular momentum component along the direction
$\mathbf{n}_{i}'$ with the spherical angles $\theta_{i}^{^{\prime}},\phi
_{i}^{^{\prime}}$. 
The spin coherent states fulfill
\begin{equation}
\left\langle {V}_{i}^{\prime\prime}{}^{\ast}|U_{i}^{\prime}\right\rangle
=\left(  1+V_{i}^{^{\prime\prime}}U_{i}^{\prime}\right)  ^{2j} \label{overlap}%
\end{equation}
and the overcompleteness relation
\begin{equation}
\label{overcomp}\frac{2j+1}{\pi}\int\frac{d^{2}U_{i}^{\prime}}{\left(
1+{U^{\prime}}_{i}^{\ast}U_{i}^{\prime}\right)  ^{2j+2}}\left\vert
U_{i}^{\prime}\right\rangle \left\langle U_{i}^{\prime}\right\vert
=\mathds{1}
\end{equation}
holds. Here $d^{2}U_{i}^{\prime}$ is a shorthand for $d\Re U_{i}^{\prime}d\Im
U_{i}^{\prime}$.

The asymptotic expression for the propagator in the semiclassical limit
$\hbar\rightarrow0,\quad J^{2}=\hbar^{2}j\left(  j+1\right)  =\mathrm{const,}
$ is derived in \cite{Garg} by slicing $t$ into small intervals. This yields a
Feynman path integral expression in which the dynamics in the
short intervals is glued together by stationary phase. Here we give the result in a slightly
reformulated version as compared to \cite{Garg0}. As derived there, the classical dynamics is
determined by the equations
\begin{equation}
\dot{U}_{i}=-\frac{i}{2J_{\mathrm{class}}}(1+U_{i}V_{i})^{2}\frac{\partial
H}{\partial V_{i}},\hspace*{6mm}\dot{V}_{i}=\frac{i}{2J_{\mathrm{class}}%
}(1+U_{i}V_{i})^{2}\frac{\partial H}{\partial U_{i}} \label{motion}%
\end{equation}
where $J_{\mathrm{class}}\equiv\hbar\left(  j+1/2\right)  \approx\sqrt
{\hbar^{2}j\left(  j+1\right)  }$. In contrast to \cite{Garg0}, the classical Hamilton function $H\left(
\mathbf{U,V}\right)  $ is obtained from the operator $\mathcal{H}%
(\mathbf{{\hat{J}}}_{1},\ldots,\mathbf{{\hat{J}}}_{N})$ by the
\textquotedblleft naive\textquotedblright\ semiclassical substitution%
\begin{equation}
\boldsymbol{{\mathcal{J}}}_{i}\rightarrow\mathbf{J}_{i,\mathrm{class}}%
=J_{\mathrm{class}}\mathbf{n}_{i}. \label{classdreh}%
\end{equation}
on which we elaborate more in the next subsection.
Here $\mathbf{n}_{i}$ stands for the unit vector with the spherical angles
$\theta_{i},\phi_{i}$. Expressing the latter in terms of $U_{i},V_{i}$
according to
\begin{eqnarray}
U_{i}  &  =\mathrm{e}^{-i\phi_{i}}\cot\frac{\theta_{i}}{2}%
,\label{uvtothetaphi}\\
V_{i}  &  =\mathrm{e}^{i\phi_{i}}\cot\frac{\theta_{i}}{2}\nonumber
\end{eqnarray}
we get the explicit substitution rules,%
\begin{eqnarray}
\mathcal{J}_{i,1}  &  \rightarrow J_{\mathrm{class}}n_{i,1}=J_{\mathrm{class}%
}\frac{U_{i}+V_{i}}{U_{i}V_{i}+1},\nonumber\\
\mathcal{J}_{i,2}  &  \rightarrow J_{\mathrm{class}}n_{i,2}=J_{\mathrm{class}%
}\frac{V_{i}-U_{i}}{i\left(  U_{i}V_{i}+1\right)  },\nonumber\\
\mathcal{J}_{i,3}  &  \rightarrow J_{\mathrm{class}}n_{i,3}=J_{\mathrm{class}%
}\frac{U_{i}V_{i}-1}{U_{i}V_{i}+1}. \label{unitvectorn}%
\end{eqnarray}
The asymptotic expression for the propagator is then
\begin{equation}
K(\mathbf{U}^{\prime},\mathbf{V}^{\prime\prime},t)=\sum_{\gamma}
{\det}^{1/2}\left(  \frac{i}{2J_{\mathrm{class}}}\frac{\partial^{2}S_{\gamma}%
}{\partial\mathbf{U}^{\prime}\partial\mathbf{V}^{\prime\prime}}\right)
\mathrm{e}^{iS_{\gamma}/\hbar}. \label{prop1}%
\end{equation}
The sum runs over
all trajectories $\gamma$ determined by Eq.\ (\ref{motion}) with the boundary
conditions $\mathbf{U}(0)=\mathbf{U}^{\prime}$ and $\mathbf{V}(t)=\mathbf{V}%
^{\prime\prime}$. The quantity $S_{\gamma}$ is the classical action
\begin{eqnarray}
S_{\gamma}(\mathbf{U}^{\prime},\mathbf{V}^{\prime\prime},t)  &  =-iJ_{\mathrm{class}%
}\sum_{i=1}^{N}\left[  \ln(1+V_{i}^{\prime\prime}U_{i}^{\prime\prime}%
)+\ln(1+V_{i}^{\prime}U_{i}^{\prime})\right] \nonumber\label{actio}\\
&  -\int_{0}^{t}dt^{\prime}\left[  iJ_{\mathrm{class}}\sum_{i=1}^{N}\frac
{\dot{V}_{i}U_{i}-V_{i}\dot{U}_{i}}{(1+U_{i}V_{i})}+H\right]
\end{eqnarray}
which solves the Hamilton Jacobi equations
\begin{equation}
i\frac{\partial S_{\gamma}}{\partial V_{i}^{\prime\prime}}=\frac{2J_{\mathrm{class}%
}U_{i}^{\prime\prime}}{1+U_{i}^{\prime\prime}V_{i}^{\prime\prime}}%
,\hspace*{6mm}i\frac{\partial S_{\gamma}}{\partial U_{i}^{\prime}}=\frac
{2J_{\mathrm{class}}V_{i}^{\prime}}{1+U_{i}^{\prime}V_{i}^{\prime}}
\label{Hamilton}%
\end{equation}
As $U_{i}$ is mostly not equal to $V_{i}^{\ast}$ the trajectory on the unit sphere $\theta
_{i}=\theta_{i}\left(  t\right)  $, $\phi_{i}=\phi_{i}\left(  t\right)  $ is, according to
(\ref{uvtothetaphi}),
complex. This is natural since arbitrarily chosen initial $\mathbf{U}^{\prime
}$ and final $\mathbf{V}^{\prime\prime}$ points are not connected by a
classical trajectory at time $t$. Indeed, the system of $N$ spins has $N$
degrees of freedom in the classical limit, implying that a real trajectory is fixed by
$2N,$ not $4N,$ real parameters. The Hamilton function and the action are then
complex and the propagator is exponentially small. On the other hand, for
classical trajectories which solve (\ref{motion}) we must have $U_{i}%
=V_{i}^{\ast}$ such that the solution of (\ref{motion}) must additionally obey
$U_{i}\left(  t\right)  =V_{i}^{\prime\prime\ast}$ and $U_{i}^{\prime}%
=V_{i}^{\ast}\left(  0\right)  .$ Here the factor of $\hbar/2$ in the 
definition of $J_{\mathrm{class}}$ becomes essential, otherwise an inadmissible error $O\left(
\hbar^{0}\right)  $ will be introduced in the exponent of the propagator.
\subsection{Solari-Kochetov phase}\label{subsec22}
In many papers on the propagator in the coherent state representation,
including \cite{Garg}, the Hamilton function is defined as
\begin{equation}\label{h00}
h\left(  \mathbf{U,V}\right)  =\frac{\left\langle \mathbf{V}^{\ast}\left\vert
\mathcal{H}\right\vert \mathbf{U}\right\rangle }{\left\langle \mathbf{V}%
^{\ast}|\mathbf{U}\right\rangle }.
\end{equation}
Compared with the \textquotedblleft naive\textquotedblright\ $H\left(
\mathbf{U,V}\right)  ,$ it contains non-classical additional terms, namely,%
\begin{equation}
h\left(  \mathbf{U,V}\right)  =H\left(  \mathbf{U,V}\right)  +\hbar Z+O\left(
\hbar^{2}\right)  , \label{hprime}%
\end{equation}
with
\begin{equation}
Z=\frac{1}{4J_{\mathrm{class}}}\sum_{i=1}^{N}(1+U_{i}V_{i})^{2}\frac
{\partial^{2}H}{\partial U_{i}\partial V_{i}}. \label{Z}%
\end{equation}
If $h\left(  \mathbf{U,V}\right)  $ is chosen as the Hamilton function the
spurious contribution of $\hbar Z$ to the classical action needs to be
compensated by the so called Solari-Kochetov (SK) correction phase
\cite{Sola},%
\begin{equation}
\Delta\Phi_{SK}=\int_{0}^{t}Zdt^{\prime}.
\end{equation}
Relations equivalent to (\ref{hprime}) formulated as the connection between
the $Q$-representation and the Weyl symbol, were established for the
translational Hamiltonians in \cite{Barang} and for the spin Hamiltonians in
\cite{Braun1,Pletyu}. 

In the remainder of this subsection we give an elementary proof of the relation (\ref{hprime}) by induction. To
simplify the notation we restrict ourselves to $N=1$. Compare the Hamilton functions
$h\left(  U,V\right)  =\left\langle V^{\ast}\left\vert \mathcal{H}\right\vert
U\right\rangle /\left\langle V^{\ast}|U\right\rangle $ and
\begin{equation}
\tilde{h}(U,V)=\frac{\left\langle V^{\ast}\left\vert \tilde{\mathcal{H}%
}\right\vert U\right\rangle }{\left\langle V^{\ast}|U\right\rangle }
\label{Hm}%
\end{equation}
with the spin Hamiltonian $\tilde{\mathcal{H}}=\left(  \mathcal{J}_{m}%
\mathcal{H}+\mathcal{H}\mathcal{J}_{m}\right)  /2$ containing an additional spin
operator $\mathcal{J}_{m}$, with fixed but arbitrary $m=1,2,3$. The relations how 
angular momentum operators act on the coherent states (\ref{cohstate}) (see \cite{Viera}, Eqs.
(4.1)-(4.3))
\begin{eqnarray*}
\mathcal{J}_{3}\left\vert U\right\rangle  &  =\hbar\left(  U\frac{d}{dU}-j\right)
\left\vert U\right\rangle ,\\
\mathcal{J}_{-}\left\vert U\right\rangle  &  =\hbar\left(  2jU-U^{2}\frac{d}%
{dU}\right)  \left\vert U\right\rangle ,\\
\mathcal{J}_{+}\left\vert U\right\rangle  &  =\hbar\frac{d}{dU}\left\vert
U\right\rangle 
\end{eqnarray*}
show that $\left\langle V^{\ast}\left\vert \tilde{\mathcal{H}}\right\vert
U\right\rangle $ can be obtained from $\left\langle V^{\ast}\left\vert
\mathcal{H}\right\vert U\right\rangle $ as a combination of its derivatives by
$U$ and $V$. Combining the result with $\left\langle V^{\ast}|U\right\rangle $
from (\ref{overlap}) we derive the exact relation between the Hamiltonians (\ref{h00}) and (\ref{Hm})
\begin{equation}
\tilde{h}=\hbar\,jn_{m}h+\frac{\hbar}{2}\mathcal{R}_{m} h  ,\quad
m=1,2,3, \label{recH}%
\end{equation}
with the differential operator 
\begin{eqnarray*}
\mathcal{R}_{1}  &  =\frac{1-U^{2}}{2}\frac{\partial}{\partial U}+\frac{1-V^{2}%
}{2}\frac{\partial}{\partial V}\\
\mathcal{R}_{2}  &  =\frac{1+U^{2}}{2i}\frac{\partial}{\partial U}-\frac{1+V^{2}%
}{2i}\frac{\partial}{\partial V},\\
\mathcal{R}\,_{3}  &  =V\frac{\partial}{\partial V}+U\frac{\partial}{\partial U}.
\end{eqnarray*}
acting on the Hamiltonian (\ref{h00}). The functions $n_{m}\left(  U,V\right)  $ are defined in Eq.\ (\ref{unitvectorn}).
Next we expand $h$ and $\tilde{h}$ in (\ref{recH}) with respect to $\hbar$ for $\hbar\rightarrow 0$
\[
h=h_{0}\left(  1+\frac{\hbar}{J_{\rm{class}}}W+\ldots\right)  ,\quad\tilde
{h}=\tilde{h}_{0}\left(  1+\frac{\hbar}{J_{\rm{class}}}\tilde{W}+\ldots\right)  ,
\]
substitute $\hbar j=J_{\rm{class}}-\frac{\hbar}{2}$ and compare terms of the same
order in $\hbar$. The zeroth order yields the \textquotedblleft
naive\textquotedblright\ classical substitution,
\begin{equation}
\tilde{h}_{0}=J_{\mathrm{class}}n_{m}h_{0} \label{hclas}%
\end{equation}
while the next order gives,
\begin{equation}
\tilde{W}=W-\frac{1}{2}+\frac{1}{2n_{m}}\mathcal{R}_{m}\ln h_{0}
,\quad m=1,2,3. \label{recurSC}%
\end{equation}
Now consider the quantities $Z$ calculated using $h_{0}$ instead of $H$ in (\ref{Z}) and analogously
$\tilde{Z}$ calculated using $\tilde{h}_{0}$. After defining $W_{SK}$ and $\tilde{W}_{SK}$ by
\begin{equation}
Z\equiv h_{0}\frac{\hbar}{J_{\mathrm{class}}}\,W_{SK},\hspace*{8mm}\tilde
{Z}\equiv\,\tilde{h}_{0}\frac{\hbar}{J_{\mathrm{class}}}\tilde{W}_{SK},
\end{equation}
the relation between $W_{SK}$ and $\tilde{W}_{SK}$ follows from Eqs.\ (\ref{Z}%
), (\ref{hclas}),
\[
\fl\tilde{W}_{SK}=W_{SK}+\frac{\left(  1+UV\right)  ^{2}}{4}\left(  \frac
{1}{n_{m}}\frac{\partial^{2}n_{m}}{\partial U\partial V}+\frac{\partial\ln
n_{m}}{\partial U}\frac{\partial\ln h_{0}}{\partial V}+\frac{\partial\ln
n_{m}}{\partial V}\frac{\partial\ln h_{0}}{\partial U}\right)  .
\]
It is easy to check that after insertion of the explicit $n_{m}$ from
Eq.\ (\ref{unitvectorn}) the last relation becomes identical to
Eq.\ (\ref{recurSC}).

Starting from $\mathcal{H}=1$ when $W=0$ and $W_{SK}=0$ trivially coincide we
can build any spin Hamiltonian by consecutive $\mathcal{H}\rightarrow\tilde{\mathcal{H}%
}$ steps each time changing $W$ and $W_{SK}$ by the same amount. Therefore, for
all spin Hamiltonians we have $W=$ $W_{SK}$ and the term $Z$ in Eq.\
(\ref{hprime}) is given by Eq. (\ref{Z}). The SK correction thus indeed cancels
the non-classical terms in $h\left(  U,V\right)  $ resulting in the
\textquotedblleft naive\textquotedblright\ Hamilton function $H\left(
U,V\right)  $. Importantly, ordering of the angular momentum operators in the
spin Hamiltonian $\mathcal{H}$ is semiclassically insignificant as long as it
is Hermitian and has real coefficients since various orderings differ then at most by $O\left(
\hbar^{2}\right)  $.

\section{Trace formula}\label{secIII}
The aim is to derive an expression in terms of a sum over classical orbits for
the leading fluctuating part $d_{\mathrm{osc}}(E)$ of the density of states starting from the propagator 
introduced in the last section. In subsection \ref{subsec31} we give the density of states as multiple integral of the propagator.
To perform these integrals within saddle point approximation we introduce canonical deviations from the saddle points 
in subsection \ref{subsec32}. In subsection \ref{subsec33} we rotate the coordinate system such that the integral becomes especially simple.
Finally, we give the resulting expression for $d_{\mathrm{osc}}(E)$ in subsection \ref{subsec34}. 
\subsection{Trace of the propagator as a periodic orbit sum}\label{subsec31}
The density of states $d(E)$ is obtained from the propagator by
\begin{equation}
d(E)=-\frac{1}{\pi}\Im\left[  \frac{1}{i\hbar}\int_{0}^{\infty
}dt\mathrm{Tr}K(t)\mathrm{e}^{iEt/\hbar}\right]  .\label{four}%
\end{equation}
The density $d(E)$ splits into a mean and an oscillating part. The mean part can be
related to orbits with zero period in (\ref{prop1}). However, we will concentrate
in this section on orbits with nonzero period and thus on the oscillating part of the density of states $d_{\rm osc}(E)$.
We begin with computing the trace of the propagator,
\begin{equation}
\mathrm{Tr}K(t)=\left(  \frac{2J_{\mathrm{class}}}{\pi\hbar}\right)  ^{N}\left.%
\int_{-\infty}^{\infty}\left(  \prod_{i=1}^{N}\frac{d^2U_{i}^{\prime}}
{\left(  1+\left\vert U_{i}^{\prime}\right\vert ^{2}\right)
^{2j+2}}\right)    K(\mathbf{U}^{\prime
},\mathbf{V}^{\prime\prime},t)\right|_{\mathbf{V}^{\prime\prime}=\left(  \mathbf{U}^{\prime
}\right)  ^{\ast}}.\label{traK}%
\end{equation}
Here $\mathbf{V}^{\prime\prime}$  is a complex conjugate of $\mathbf{U}%
^{\prime}$, as the trajectories are closed. Inserting the
expression (\ref{prop1}) in the last equation, we obtain a sum over closed
classical trajectories $\gamma$ from $\mathbf{U}^{\prime}$ to $\left(
\mathbf{V}^{\prime\prime}\right)  ^{\ast}=\mathbf{U}^{\prime}$ with duration
$t$. Concentrating on one element of the sum, which we write as
\begin{eqnarray}
\left[  \mathrm{Tr}K(t)\right]  _{\gamma}=\left(  \frac{i\,2J_{\mathrm{class}%
}}{\pi^{2}\hbar^{2}}\right)  ^{N/2}\int_{-\infty}^{\infty}\left(  \prod
_{i=1}^{N}\frac{d^2{U}_{i}^{\prime}}{1+\left\vert
U_{i}^{\prime}\right\vert ^{2}}\right) \nonumber\\\times \left.  {\det}^{1/2}\left(
\frac{\partial^{2}S_{\gamma}}{\partial\mathbf{U}^{\prime}\partial
\mathbf{V}^{\prime\prime}}\right)\mathrm{e}^{iF_{\gamma
}/\hbar}\right\vert _{\mathbf{V}^{\prime\prime}=\left(  \mathbf{U}^{\prime
}\right)  ^{\ast}},\label{TraKgamma}%
\end{eqnarray}
we perform the integrals via saddle point approximation by computing the
stationary points of
\begin{equation}
F_{\gamma}(\mathbf{U}^{\prime},\mathbf{V}^{\prime\prime},t)\equiv S_{\gamma
}(\mathbf{U}^{\prime},\mathbf{V}^{\prime\prime},t)+2iJ_{\mathrm{class}}%
\sum_{i=1}^{N}\ln(1+U_{i}^{\prime}V_{i}^{\prime\prime}).
\end{equation}
Together with Eq.\ (\ref{Hamilton}) we find the conditions $V_{i}^{\prime
}=V_{i}^{\prime\prime}$ and $U_{i}^{\prime}=U_{i}^{\prime\prime}$, i.e.\ a
periodic orbit with the coordinates $\mathbf{U}^{\prime}=\mathbf{s}$ and
$\mathbf{V}^{\prime\prime}=\mathbf{s}^{\ast}$. Below the subscript $\gamma$ of
$F$ is suppressed. We note that on the periodic orbit $\gamma$ all logarithms in
$F(\mathbf{s},\mathbf{s}^{\ast},t)$ cancel and $F(\mathbf{s}%
,\mathbf{s}^{\ast},t)$ does not depend on the concrete choice of the initial and final point $\mathbf{s}$
along $\gamma$. Thus, on $\gamma$ 
\begin{equation}
S(t)=-\int_{0}^{t}dt^{\prime}\left[  iJ_{\mathrm{class}}\sum_{i=1}^{N}%
\frac{\dot{V}_{i}U_{i}-V_{i}\dot{U}_{i}}{(1+U_{i}V_{i})}+H\right]
\label{peract}%
\end{equation}
equals $F({\bf s},{\bf s}^*,t)$.
\subsection{Canonical variables for small deviations}\label{subsec32}
In order to perform the integrals in Eq.\ (\ref{TraKgamma}) by saddle point approximation, we consider 
the motion in the vicinity of the saddle point $\mathbf{s}$ described by small
deviations $\mathbf{\delta U}=\mathbf{U}-\mathbf{s}$ and $\mathbf{\delta
V}=\mathbf{\delta U}^*=\mathbf{V}-\mathbf{s}^{\ast}$. However, we prefer to use deviations that 
fulfill canonical equations, but are no longer complex conjugate to each other. As shown in Appendix A, they are given by 
\[
\delta\mathbf{\tilde{U}}=B\mathbf{\delta U},\quad\delta\mathbf{\tilde{V}%
}=\mathbf{\delta V}%
\]
where $B$ is the diagonal matrix
\[
B=\mathrm{diag}\left(  \frac{2iJ_{\mathrm{class}}}{\left(  1+|s_{1}%
|^{2}\right)  ^{2}},\ldots,\frac{2iJ_{\mathrm{class}}}{\left(  1+|s_{N}%
|^{2}\right)  ^{2}}\right).
\]
Next we introduce the $2N$-component vector $\delta\mathbf{\tilde{v}}%
=(\delta\mathbf{\tilde{U}}^{\prime},\delta\mathbf{\tilde{V}}^{\prime\prime})$, 
where primes again refer to initial and double primes to final deviations,
express  $F(\mathbf{U}^{\prime},\mathbf{V}^{\prime\prime},t)$
in terms of $\delta\mathbf{\tilde{U}}^{\prime},\delta\mathbf{\tilde{V}}^{\prime\prime}%
$ and expand around the point $\mathbf{s}$ of the periodic orbit up
to second order
\begin{equation}
F(\mathbf{U}^{\prime},\mathbf{V}^{\prime\prime},t)\approx S(t)+\frac{1}%
{2}\delta\mathbf{\tilde{v}}^{T}H_{F}\delta\mathbf{\tilde{v}.}%
\end{equation}
Here $H_{F}$ is the $2N\times2N$ Hessian matrix containing the second
derivatives of $F(\mathbf{s}+B^{-1}\delta\mathbf{\tilde{U}}^{\prime},\mathbf{s}^{\ast}+\delta\mathbf{\tilde{V}}^{\prime\prime
},t)$ with respect to
$\delta\mathbf{\tilde{U}}^{\prime}$ and $\delta\mathbf{\tilde{V}}%
^{\prime\prime}.$ The analogous expansion can be done for the action
$S(\mathbf{U}^{\prime},\mathbf{V}^{\prime\prime},t)$. Between the Hessians of
$F$ and $S$ the relation
\begin{equation}
H_{F}=H_{S}+\left(
\begin{array}
[c]{cc}%
A & \mathds{1}_{N}\\
\mathds{1}_{N} & D
\end{array}
\right)
\end{equation}
holds with the $N\times N$ diagonal matrices 
\begin{equation}
D  =-\mathrm{diag}\left(  \frac{2iJ_{\mathrm{class}}s_{1}^{2}}{\left(
1+|s_{1}|^{2}\right)  ^{2}},\ldots,\frac{2iJ_{\mathrm{class}}s_{N}^{2}%
}{\left(  1+|s_{N}|^{2}\right)  ^{2}}\right)  ,\quad
A   =-B^{-2}D^{\ast}.
\end{equation}
The complex $2N\times2N$ monodromy matrix $M$ consisting of the four $N\times N$ 
blocks $M_{aa}$, $M_{ab}$, $M_{ba}$ and $M_{bb}$ 
relates the initial and final deviations,
\begin{eqnarray}\label{monodrom}
\left(
\begin{array}
[c]{c}%
\delta\mathbf{\tilde{U}}^{\prime\prime}\\
\delta\mathbf{\tilde{V}}^{\prime\prime}%
\end{array}
\right)  =\left(
\begin{array}
[c]{cc}%
M_{aa} & M_{ab}\\
M_{ba} & M_{bb}%
\end{array}
\right)  \left(
\begin{array}
[c]{c}%
\delta\mathbf{\tilde{U}}^{\prime}\\
\delta\mathbf{\tilde{V}}^{\prime}%
\end{array}
\right)  ;
\end{eqnarray}
it is symplectic because the deviations obey canonical equations.  Variation
of (\ref{Hamilton}) with respect to $\mathbf{U}^{\prime}$, $\mathbf{V}%
^{\prime}$, $\mathbf{U}^{\prime\prime}$ and $\mathbf{V}^{\prime\prime}$ yields
\begin{equation}
\left[  H_{S}+\left(
\begin{array}
[c]{cc}%
A & 0\\
0 & D
\end{array}
\right)  \right]  \left(
\begin{array}
[c]{c}%
\delta\mathbf{\tilde{U}}^{\prime}\\
\delta\mathbf{\tilde{V}^{\prime\prime}}%
\end{array}
\right)  =-\left(
\begin{array}
[c]{cc}%
0 & \mathds{1}_{N}\\
\mathds{1}_{N} & 0
\end{array}
\right)  \left(
\begin{array}
[c]{c}%
\delta\mathbf{\tilde{U}}^{\prime\prime}\\
\delta\mathbf{\tilde{V}^{\prime}}%
\end{array}
\right),  \label{grads}%
\end{equation}
see the analogous relation for the non-canonical variables in \cite{Ribeiro}%
. By partially inverting (\ref{monodrom}) such that it takes the form
(\ref{grads}), $H_{S}$ is expressed in terms of the monodromy matrix
\begin{equation}
H_{S}=\left(
\begin{array}
[c]{cc}%
M_{bb}^{-1}M_{ba}-A & -M_{bb}^{-1}\\
M_{ab}M_{bb}^{-1}M_{ba}-M_{aa} &
-M_{ab}M_{bb}^{-1}-D
\end{array}
\right)  \label{HS}%
\end{equation}
yielding for $H_{F}$
\begin{equation}
H_{F}=\left(\begin{array}{cc}
             H_{aa}&H_{ab}\\H_{ba}&H_{bb}
            \end{array}
\right)=\left(
\begin{array}
[c]{cc}%
M_{bb}^{-1}M_{ba} & \mathds{1}_{N}-M_{bb}^{-1}\\
\mathds{1}_{N}+M_{ab}M_{bb}^{-1}M_{ba}-M_{aa} &
-M_{ab}M_{bb}^{-1}%
\end{array}
\right)  .\label{hftom}%
\end{equation}
The symmetry of the Hessians can be checked using
the symplectic property of $M$.  In the sequel,  ``the three-determinants-identity'' 
\begin{equation}\label{3dets}
 \det M_{bb}\det H_F=\det(M-\mathds{1}_{2N})
\end{equation} 
will be important. It is valid for any symplectic matrix $M$ with the subblock $M_{bb}$ and the symmetric matrix $H_F(M)$ connected with $M$ via (\ref{hftom}). 
Although we can hardly believe that it is new, we could not find it in the literature and give the proof in Appendix B.

Equation (\ref{HS}) shows that the
matrix of the second mixed derivatives of the action and its determinant are related to the
monodromy matrix as,%
\begin{eqnarray}\label{trafo22}
\frac{\partial^{2}S}{\partial\mathbf{U}^{\prime}\partial\mathbf{V}%
^{\prime\prime}}=B\frac{\partial^{2}S}{\partial\delta\mathbf{\tilde{U}%
}^{\prime}\partial\delta\mathbf{\tilde{V}}^{\prime\prime}}=-BM_{bb}^{-1},\nonumber\\
\det\frac{\partial^{2}S}{\partial\mathbf{U}^{\prime}\partial\mathbf{V}%
^{\prime\prime}}=\frac{\det B}{\det(-M_{bb})}.
\end{eqnarray}
The contribution to the integral Eq.\ (\ref{TraKgamma}) of the vicinity of the
point $\mathbf{s}$ can be calculated using the real and imaginary parts of
$\delta\mathbf{\tilde{U}}^{^{\prime}}$ as the integration variables with the
Jacobian $1/\det B$.

Taking into account that 
\begin{equation}
 \prod_{i=1}^N\frac{1}{1+|U_i'|^2}\approx\sqrt{\frac{\det B}{\left(2iJ_{\rm class}\right)^N}}
\end{equation}
and collecting all $B$-dependent terms in Eq.\ (\ref{TraKgamma}) we
obtain from Eq.\ (\ref{TraKgamma}),
\begin{eqnarray}\label{haha}
 \left[{\rm Tr}K(t)\right]_\gamma&=&\left(\frac{1}{\pi\hbar}\right)^N{\det}^{-1/2}(-M
_{bb}){\rm e}^{iS(t)/\hbar}\nonumber\\&&\times
\int_{-\infty}^\infty\prod_{i=1}^N d^2\delta\tilde U_i' \frac{1}{ \det B}\,
{\rm e}^{i\delta\tilde{\bf v}^TH_F\delta\tilde{\bf v}/2\hbar}.
\end{eqnarray}
The action $S({\bf U}',{\bf V}'',t)$ does not change under the shift along the orbit and as a result the monodromy matrix has a double degenerate eigenvalue $1$ while
$H_F$ possesses a zero eigenvalue (see (\ref{3dets})). To get rid of it, we slightly perturb the problem, say,  replacing $H_F$ for the moment  by 
$H_F^\epsilon=H_F+\epsilon\mathds{1}_{2N}$ with $\epsilon>0$ and take the limit 
$\epsilon\rightarrow0$ at the end. By inverting relation (\ref{hftom}) 
\begin{equation}
 M=\left(\begin{array}{cc}
  \mathds{1}_N-H_{ab}^T-H_{bb}ZH_{aa}&-H_{bb}Z\\
  ZH_{aa}&Z
                   \end{array}
\right)
\end{equation}
with $Z\equiv(\mathds{1}_N-H_{ab})^{-1}$, we understand that this change preserves the symplectic 
property of the corresponding monodromy matrix $M^\epsilon$. The eigenvalue 1 is replaced in  $M^\epsilon$ by a doublet with the splitting $\propto \sqrt{\epsilon}$.

Now the Gaussian integral in Eq.\ (\ref{haha}) can be performed. Remembering that  the two parts of the vector $\delta\tilde{\bf v} $ are related as $\delta\mathbf{\tilde{V''}}=B^{-1}\delta\mathbf{\tilde{U'}}$ we have,
\begin{eqnarray}
\int_{-\infty}^\infty \prod_{i=1}^N d^2\delta\tilde U_i'
{\rm e}^{i\delta\tilde{\bf v}^TH_F^\epsilon\delta\tilde{\bf v}/2\hbar}=\frac{\pi^N\det B}{(-i)^N\sqrt{\det H_F^\epsilon}}\,.
\end{eqnarray}
The integral  (\ref{haha}) is thus obtained as $\mathrm{const.}\left[ \det\left( -M_{bb}^\epsilon\right) \det H_F^\epsilon\right]^{-1/2}$.

\subsection{Generalized eigensystem of the monodromy matrix}\label{subsec33}
At the current stage it is unclear how to perform the limit $\epsilon\rightarrow0$, therefore we 
introduce a system of variables for the 
deviations where $M$ is almost diagonal.
We consider a linear symplectic transformation $\delta{\tilde{\mathbf{v}%
}}=W\delta{\mathbf{x}}$ introducing $2N$ real canonical variables
$\delta{\mathbf{x}}=\left(  \mathbf{\delta q},\mathbf{\delta
p}\right)^{T}$ with the canonically conjugate
\begin{equation}
\mathbf{\delta q=}\left(  \delta t,\delta\boldsymbol{\xi}\right)  ^{T}%
,\quad\mathbf{\delta p=}\left(  \delta E,\delta\boldsymbol{\pi}\right)
^{T}\,.\label{newvar}%
\end{equation}
Here $\delta t$ is the time variation defining the shift along the direction
of motion, $\delta E$ the variation of energy leading to the shift to a point of an
infinitely close periodic orbit in the energy shell $H=E+\delta E$; the
$N-1$-componental $\delta\boldsymbol{\xi}$ and $\delta\boldsymbol{\pi}$ denote the
shifts in the direction of the stable and unstable eigenvectors of the
monodromy matrix corresponding to the eigenvalues $\Lambda_{n}$ with
$\left\vert \Lambda_{n}\right\vert <1,$ and $1/\Lambda_{n}$, respectively,
$n=1,\ldots,N-1$. For simplicity we assume that the periodic orbit is
unstable. In these coordinates the monodromy matrix reads
\cite{Ribeiro},
\begin{equation}
m=\left(
\begin{array}
[c]{cc}%
m_{aa} & m_{ab}\\
\mathbf{0}_{N} & m_{bb}%
\end{array}
\right),
\end{equation}
where the $N\times N$ blocks are diagonal matrices,
\begin{eqnarray*}
m_{aa} &  =\mathrm{diag}\left(  1,\Lambda_{1},\ldots,\Lambda_{N-1}\right)  ,\\
m_{bb} &  =\mathrm{diag}\left(  1,1/\Lambda_{1},\ldots,1/\Lambda_{N-1}\right)
,\\
m_{ab} &  =\mathrm{diag}\left(  -k,0,\ldots,0\right)  .
\end{eqnarray*}
It has a single non-zero off-diagonal element $-k$. One of the two eigenvalues
equal to unity is associated with $\delta t$ and corresponds to the
eigenvector tangent to the orbit in the phase space. The other one is
connected with $\delta E$; in the assumption $k\neq0$, it corresponds not to an eigenvector
but to an associated eigenvector of the matrix $m$ which is then not diagonalizable.

To find $k$ we consider the change $T\rightarrow T+\delta T$ of the orbit period
caused by the change of energy. After the time $T$ the point traveling along
the infinitely close periodic orbit with the energy $E+\delta E$ will either
not yet return to the initial position if $\delta T>0$, or overrun it if
$\delta T<0$. Consequently after the time $T$ the energy shift $\delta E$
leads to the shift along the orbit described by $\delta t=-\delta T=-\left(
dT/dE\right)  \delta E$ such that \cite{Ribeiro,Barang}%
\[
k=\frac{dT}{dE}.
\]
The transformed Hessian $h_F\equiv H_F(m)$
consists of the subblocks,
\begin{eqnarray*}
\left(  h_{F}\right)  _{aa}=\mathbf{0}_{N},\qquad\left(  h_{F}\right)  _{bb}  =\mathrm{diag}\left(
k,0,\ldots,0\right)  ,\\
\left(  h_{F}\right)  _{ab}   =\left(  h_{F}%
\right)  _{ba}=\mathrm{diag}\left(  0,1
-\Lambda
_{1},\ldots,1-\Lambda
_{N-1}\right) 
 .\end{eqnarray*}
 Let  $h_F^\epsilon$ be the Hessian resulting from the monodromy matrix $m^\epsilon=WM^\epsilon W^{-1}$
converging to $h_{F}$ for $\epsilon\rightarrow0$. According to the three-determinants-relation (\ref{3dets}) the product $\left[\det(-M_{bb})\det H_F^\epsilon\right]^{-1/2})$ is canonically invariant and can be replaced by 
$\left[\det(-m_{bb}^\epsilon)\det h_F^\epsilon\right]^{-1/2}$. 
Next the factor $\left(\det h_F^{\epsilon}\right)^{-1/2}$ can be replaced by a Gaussian integral in the coordinates 
$\delta{\bf x}$. 
Now the limit $\epsilon\rightarrow0$ can be performed. Collecting the
prefactors we obtain,
\begin{equation}
\fl\left[  \mathrm{Tr}K(t)\right]  _{\gamma}\sim\left(  \frac{1}{i\pi\hbar
}\right)  ^{N} \frac{{\rm e}^{iS_{\gamma}/\hbar}}{{\det}^{1/2} (-m_{bb})}\int_0^{t_{\gamma}^{P}} d\delta
t\int_{-\infty}^\infty d\delta Ed\delta\boldsymbol{\xi}d\delta\boldsymbol{\pi}\exp\frac
{i\delta{{\mathbf{x}}^{T}}h_{F}\delta{\mathbf{x}%
}}{2\hbar}.\label{eq222}
\end{equation}
The inner integrals are Gaussian, and the result of integration does not
depend on the position $\mathbf{s}$ on the orbit, i. e., on the time variable.
Therefore the integral with respect to $\delta t$ yields the primitive period
of the orbit $t_{\gamma}^{P}=t/r_{\gamma},$ where $r_{\gamma}$ stands for the
number of repetitions of the orbit $\gamma$. The Gaussian integral with respect 
to $\delta E$ brings about the factor $1/\sqrt{k}$. 
\subsection{Semiclassical expressions for the trace of the propagator and the density of states}\label{subsec34}
Finally
we obtain for the trace of
the propagator in the semiclassical limit
\begin{equation}
\mathrm{Tr}K(t)  \sim\frac{1}{\sqrt{-2i\pi\hbar k}}\sum_{\gamma%
}\frac{t_{\gamma}^{P}}{\sqrt{\left\vert \det(m_{\gamma%
}^{\mathrm{red}}-\mathds{1})\right\vert }}\label{trpro}
  \exp\left[  i\left(  \frac{S_{\gamma}(t)}{\hbar}+G_{\gamma%
}(t)\right)  \right]  .
\end{equation}
with the sum running over all periodic orbits $\gamma$ of duration $t$. Here the
reduced monodromy matrix $m_{\gamma}^{\mathrm{red}}$ is obtained from
the matrix $m_\gamma$ of the orbit $\gamma$ by omitting the directions
related to variations of $\delta t$ and $\delta E$. It is obtained from combining the
result of the Gaussian integrals in Eq.\ (\ref{eq222}) with respect to $\boldsymbol{\xi}$ and $\boldsymbol{\pi}$ with the prefactor ${\det}^{-1/2} (-m_{bb})$.
The symbol $G_{\gamma
}(t)$ stands for the Maslov phase resulting from the saddle point integrations.

The final Fourier transform from the time to the energy domain in
Eq.\ (\ref{four}) yields the Gutzwiller sum over periodic orbits $\gamma'$ at energy $E$
which extends the result of \cite{Ribeiro} to several interacting spins,
\begin{equation}
d_{\mathrm{osc}}(E)\sim\frac{1}{\pi\hbar}\sum_{\gamma'}%
\frac{t_{\gamma'}^{P}}{\sqrt{\left\vert \det(m_{\gamma'%
}^{\mathrm{red}}-\mathds{1})\right\vert }}\cos\left[  \frac{\mathcal{S}%
_{\gamma'}(E)}{\hbar}+\mathcal{G}_{\gamma'}(E)\right]  \label{tracfor}%
\end{equation}
where $\mathcal{S}_{\gamma'}(E)$ is the Legendre transform of the
action $S_{\gamma'}(t)$ and $\mathcal{G}_{\gamma'}(E)$ the Maslov phase.


\section{Trace formula for periodically driven systems}\label{secIV}

The dynamics in periodically driven systems is governed by an explicitly time
dependent Hamiltonian, thus the energy is no longer conserved. The conserved
quantity in the quantum system is in this case the quasienergy, i.e.\ the
eigenphase of the Floquet (one period time evolution) operator. Prominent
examples of periodically driven systems are kicked maps like the kicked rotor
or the kicked top in the one particle domain, see for example \cite{Haake} for
an overview, or the kicked Ising chain \cite{Prosen,Pineda} that serves as a model
system to simulate effects in many particle systems.

To obtain the trace formula for the density of eigenphases of the quantum
system, one starts from the time evolution operator for the period of driving
$t_{0}$. Its semiclassical expression is again of the form
(\ref{prop1}). In contrast to the last section the monodromy matrix for a
fully chaotic system does not possess eigenvalues equal to one as in general energy 
is not conserved and a $\delta t$-translation does not leave the orbit invariant. 
Thus, no special action to separate these directions is needed when evaluating
the Gaussian integral resulting from the saddle point approximation in
(\ref{TraKgamma}).

Performing this integral and combining the result with the prefactor of the
exponential in (\ref{prop1}) and the factor $\prod_{i=1}^{N}\left(
1+|U_{i}^{\prime}|^{2}\right) ^{-1}$ resulting from the trace integrals, we
obtain for the trace of the propagator at time $t=nt_{0}$ with $n\in
\mathds{N}$ in the limit $\hbar\rightarrow0$
\begin{equation}
\label{tratrapro}\mathrm{Tr} K(t)\sim\sum_{\gamma}\frac{n^{P}%
_{\gamma}}{\sqrt{\left|  \det\left(  M_{\gamma}-1\right)
\right|  }}\exp\left[  i\left(  \frac{S_{\gamma}(t)}{\hbar}%
+G_{\gamma}(t)\right)  \right]
\end{equation}
with the sum running over all periodic orbits $\gamma$ of duration $t$. Here
$n^{P}_{\gamma}\in\mathds{N}$ is the discrete primitive period of the orbit, that
means $n^{P}_{\gamma}=n/r_{\gamma}$ for an orbit that repeats
a shorter periodic orbit $r$ times and $M_{\gamma}$ the $2N\times
2N$-dimensional monodromy matrix for a system consisting of $N$ spins. 
Other quantities in the last equation have the same meaning as in
Eq.\ (\ref{trpro}). We note that the expression given in \cite{Haake} for
particles is of the same form as the one in Eq.\ (\ref{tratrapro}) for spins.
Given that the variables $U_{i},V_{i}$ are not canonically conjugate
and that the trace (\ref{overcomp}) involves an additional
factor $\prod_{i=1}^{N}\left(  1+|U_{i}^{\prime}|^{2}\right)  ^{-2j-2}$, this
is a nontrivial result.

In order to obtain a periodic orbit expansion for the density of eigenphases
$\rho(\theta)$, the resulting expression for the trace (\ref{tratrapro}) needs
to be inserted in the following expression for the eigenphases $\theta_{n}$
for a $N$ particle spin system \cite{Haake}
\begin{equation}
\rho(\theta)=\frac{1}{(2\tilde{j})^{N}}\sum_{n=1}^{(2\tilde{j})^{N}}%
\delta(\theta-\theta_{n})=\frac{1}{2\pi}+\frac{1}{(2\tilde{j})^{N}\pi
}\mathrm{Re}\sum_{n=1}^{\infty}\mathrm{e}^{in\theta}\mathrm{Tr}K(nt_{0}).
\end{equation}

\section{Conclusions}\label{secV}

We studied a spin chain consisting of $N$ spin-$j$ particles. 
Currently there is a considerable interest in such systems both from theoretical and from
experimental side. The goal is to investigate the properties of many
particle systems, which are still small enough that finite size effects
are important. For such a system we derived a trace formula connecting the
classical dynamics obtained in the limit of large $j$ and the quantum energy levels for time-continuous as well as for periodically driven dynamics.

Similar to the trace formula for particles it can be used to predict the properties
of the quantum spectra of the spin chains following from their classical dynamics.
It can work also in the opposite direction and give information on the particular
classical orbits responsible for the non-universal features of the quantum spectra.
We restricted ourselves to distinguishable spins but it would be interesting as
well to consider the non-distinguishable ones and their coupling to the bosonic
particles.
\appendix
\section{Canonical Variables}
We consider deviations $\delta{\bf U}$ and $\delta{\bf V}$ from a trajectory obtained as a solution of the equation of motion (\ref{motion}). Such deviations fulfill up to linear order in the deviations
the equations
\begin{eqnarray}
\delta\dot{U}_i&=&\frac{\partial H}{\partial V_i}\left(\frac{\partial k_i}{\partial U_i}\delta U_i+\frac{\partial k_i}{\partial V_i}\delta V_i\right)+k_i\left(\frac{\partial^2 H}
{\partial V_i\partial V_k}\delta V_k+\frac{\partial^2 H}{\partial U_k\partial V_i}\delta U_k\right)\nonumber\\
\delta\dot{V}_i&=&-\frac{\partial H}{\partial U_i}\left(\frac{\partial k_i}{\partial U_i}\delta U_i+\frac{\partial k_i}{\partial V_i}\delta V_i\right)-k_i\left(\frac{\partial^2 H}
{\partial U_i\partial U_k}\delta U_k+\frac{\partial^2 H}{\partial U_i\partial V_k}\delta V_k\right)
\end{eqnarray}
with
\begin{equation}
k_i=\frac{\left(1+U_{i}V_{i}\right)^2}{2iJ_{\rm class}}
\end{equation}
and the unperturbed trajectory determined by the complex coordinates $U_{i}$ and $V_{i}$ for the $i$-th spin.
The aim of this appendix is to show that the variables $\delta\tilde{\bf U}$ and $\delta\tilde{\bf V}$ defined by
\begin{equation}
\delta U_i= k_i\delta\tilde{U}_i,\hspace*{1cm}\delta V_i=\delta\tilde{V}_i
\end{equation}
are canonical. This can be done by noting that they fulfill the canonical equations of motion
\begin{equation}
\delta\dot{\tilde{U}}_i=\frac{\partial\tilde{H}}{\partial\delta\tilde{V}_i},\hspace*{1cm}
\delta\dot{\tilde{V}}_i=-\frac{\partial\tilde{H}}{\partial\delta\tilde{U}_i}
\end{equation}
with the Hamiltonian $\tilde{H}\left(\delta{\bf \tilde U},\delta{\bf \tilde V}\right)$
\begin{eqnarray}
\tilde{H}&=&\sum_{k,j=1}^N\left(\frac{1}{2}\frac{\partial^2H}{\partial U_k\partial U_j}k_kk_j\delta \tilde{U}_k\delta \tilde{U}_j
+\frac{\partial^2H}{\partial U_j\partial V_k}k_j\delta \tilde{U}_j\delta \tilde{V}_k+\frac{1}{2}\frac{\partial^2H}
{\partial V_k\partial V_j}\delta \tilde{V}_k\delta \tilde{V}_j\right)\nonumber\\&&\!\!\!\!\!\!+\sum_{j=1}^N\frac{1}{1+U_{j}V_{j}}
\left(\frac{\partial H}{\partial V_j}U_{j}\delta \tilde{V}_j^2+2\frac{\partial H}{\partial U_j}U_{j}k_j\delta \tilde{U}_j\delta \tilde{V}_j+\frac{\partial H}{\partial U_j}V_{j}k_j^2\delta \tilde{U}_j^2\right).
\end{eqnarray}

\section{The three-determinants-relation}
Here we give the proof of the identity (\ref{3dets}). Its left hand side can be obtained as determinant of the matrix 
\begin{equation}\label{eqB1}
 H_F\left(\begin{array}{cc}\mathds{1}_N&0\\ 0&M_{bb}
                      \end{array}
\right)=\left(\begin{array}{cc}M_{bb}^{-1}M_{ba}&M_{bb}-\mathds{1}_N\\ \mathds{1}_N-M_{aa}+M_{ab}M_{bb}^{-1}M_{ba}&-M_{ab}
                      \end{array}
\right)\,.
\end{equation}
The right hand side of (\ref{3dets}) can be written as
\begin{eqnarray}\label{eqB2}
 \det\left(M-\mathds{1}_{2N}\right)=\det\left(\begin{array}{cc}
                      M_{ba}&M_{bb}-\mathds{1}_N\\ \mathds{1}_N-M_{aa}&-M_{ab}
                     \end{array}
\right).
\end{eqnarray}
The equality of (\ref{eqB2}) and the determinant of (\ref{eqB1}) becomes evident when adding the 
second column of Eq.\ (\ref{eqB1}) multiplied on the right by $M_{bb}^{-1}M_{ba}$ to the first column. 

A direct consequence of the relation (\ref{3dets}) is that $\det M_{bb}\det H_F$
is invariant under a symplectic transformation $M\rightarrow WMW^{-1}$.

 \end{document}